\documentclass[showpacs,prd,twocolumn,floatfix,
nofootinbib,superscriptaddress]{revtex4}
\usepackage{graphicx}

\newcommand{\be}{\begin{equation}}
\newcommand{\ee}{\end{equation}}
\newcommand{\bea}{\begin{eqnarray}}
\newcommand{\eea}{\end{eqnarray}}

\newcommand{\psibar}{\overline{\Psi}}
\newcommand{\ubar}{\overline{u}_Q}
\newcommand{\vbar}{\overline{v}_Q}

\begin{document}


\title{ Four Fermion Operator Matching with NRQCD Heavy and 
AsqTad Light Quarks}

\author{Elvira G\'amiz}
\affiliation{Department of Physics, University of Illinois, Urbana,
 IL 61801, USA }
\author{Junko Shigemitsu}
\affiliation{Department of Physics,
The Ohio State University, Columbus, OH 43210, USA }
\author{Howard Trottier}
\affiliation{Physics Department, Simon Fraser University,
Vancouver, British Columbia, Canada}

\collaboration{HPQCD Collaboration}
\noaffiliation



\begin{abstract}
We present one-loop matching coefficients between continuum and lattice QCD 
 for the heavy-light four-fermion 
operators relevant for neutral $B$ meson mixing both within and beyond 
the Standard Model.  For the lattice theory we use nonrelativistic 
QCD (NRQCD) to describe $b$ quarks and improved staggered fermions 
(AsqTad) for light quarks.   The gauge action is the tree-level Symanzik 
improved gauge action.  Matching to full QCD is carried out 
through order $\alpha_s$, $\Lambda_{QCD}/M_b$, and $\alpha_s/(aM_b)$.  

\end{abstract}


\maketitle


\section{ Introduction}
$B^0_q - \overline{B}^0_q$ mixing, with $q=d$ or $q=s$, has been the focus 
of much attention by experimentalists and theorists in recent years,   
and it will continue to be probed extensively in the LHC era 
\cite{fleischer}.  
On the experimental front the mass differences between the ``heavy'' 
and ``light'' eigenstates in the $B^0$ systems, $\Delta  M_d$ and 
$\Delta  M_s$, are now known very accurately \cite{d01,cdf1}. Measurements 
of the decay width difference $\Delta  \Gamma_s$ and the 
phase $\phi_s$ by $D\O$ and $CDF$ have also appeared \cite{d02,d03,cdf2}. 
On the theory side neutral $B$ systems 
are of particular interest as a possible window into New Physics ($NP$). 
In the Standard Model 
 $B^0_q - \overline{B}^0_q$ mixing 
does not occur at tree level and must go through box diagrams 
involving the exchange of two $W$'s at lowest order.  
$NP$ could enter through the exchange of new particles in the box diagrams, 
or through new tree level contributions. Studies of the neutral $B$ 
meson parameters can impose important constraints on different 
NP scenarios \cite{fleischer}.

Theoretical estimates of mixing rates employ effective Hamiltonians 
involving four fermion operators.  Matrix elements of these operators 
between  $B^0_q$ and $\overline{B}^0_q$ states are needed to complete the 
calculations and this requires control over non-perturbative QCD.  
For instance 
the Standard Model expression for the mass difference $\Delta M_q$
is given by \cite{buras},
\be
\label{deltams}
\Delta M_q = \frac {G_F^2 M_W^2}{6 \pi^2} |V^*_{tq}V_{tb}|^2 \eta_2^B
S_0(x_t) M_{B_q} f^2_{B_q} \hat{B}_{B_q},
\ee
where $x_t = m_t^2/M_W^2$,  $\eta_2^B$ is  a perturbative QCD correction
factor,  $S_0(x_t)$ the Inami-Lim function and $V_{tq}$ and $V_{tb}$
 the appropriate Cabibbo-Kobayashi-Maskawa (CKM) matrix elements.
  The nonperturbative QCD
input into this formula is the combination $f^2_{B_q} \, \hat{B}_{B_q}$ 
where $f_{B_q}$ is the $B_q$ meson decay constant and 
$\hat{B}_{B_q}$ the renormalization group invariant bag parameter.
Lattice QCD provides a first principles approach to obtaining these 
crucial non-perturbative factors. The first realistic lattice 
results that include effects from two very light and the strange 
sea quarks (so-called $N_f = 2+1$ simulations) have now appeared 
\cite{ourbsmix,todd,elvira}. 

An important step in all lattice calculations is the matching 
between four-fermion operators in continuum QCD that 
enter into formulas such as eq.(\ref{deltams}) and the operators 
of the lattice theory used in the lattice QCD simulations.  In this article 
we present one loop matching results for a complete basis of $\Delta B =2$ 
four fermion operators relevant for mixing both within and beyond the 
Standard Model. In the lattice theory 
we employ a nonrelativistic QCD (NRQCD) action for the $b$ quarks,
 an improved staggered quark action (AsqTad) for the $s$ and $d$ 
quarks and the Symanzik improved glue action.

A subset of these results was used already in reference 
\cite{ourbsmix} in the calculation of all the hadronic matrix elements 
relevant for the determination of $\Delta M_s$ and $\Delta \Gamma_s$ in the 
Standard Model. That was the first lattice determination of these 
parameters with $N_f=2+1$ sea quarks.


\section{ The Four Fermion Operators and Matrix Elements in QCD}
In order to study neutral $B$ meson mixing phenomena, we focus on the 
following five $\Delta B = 2$ four-fermion operators ($i$ and $j$ are color 
indices) 
\begin{eqnarray}
\label{ffop1}
Q1 &=& \left ( \psibar^i_b \gamma^\nu P_L \Psi^i_q \right ) 
 \left ( \psibar^j_b \gamma_\nu P_L \Psi^j_q \right ) \\
\label{ffop2}
Q2 &=& \left ( \psibar^i_b  P_L \Psi^i_q \right ) 
 \left ( \psibar^j_b  P_L \Psi^j_q \right ) \\
\label{ffop3}
Q3 &=& \left ( \psibar^i_b  P_L \Psi^j_q \right ) 
 \left ( \psibar^j_b  P_L \Psi^i_q \right ) \\
\label{ffop4}
Q4 &=& \left ( \psibar^i_b P_L \Psi^i_q \right ) 
 \left ( \psibar^j_b  P_R \Psi^j_q \right ) \\
\label{ffop5}
Q5 &=& \left ( \psibar^i_b  P_L \Psi^j_q \right ) 
 \left ( \psibar^j_b  P_R \Psi^i_q \right )\, . 
\end{eqnarray}
The subscript $q$ stands for either the $d$ or the $s$ quark, both 
of which we take to be massless in our matching calculations, and 
$P_{R,L} \equiv (I \pm \gamma_5)$. 

Operators $Q1$, $Q2$ and $Q3$ appear 
in the Standard Model and are relevant for the mass and width differences 
$\Delta \, M_q$ and $ \Delta \, \Gamma_q$.  Matrix elements of $Q1$, for 
instance, lead to $f_{B_q}^2 \, B_{B_q}$.
Additional $\Delta \, B=2$ 
operators, such as $Q4$ and $Q5$
 are required when going to extensions of the Standard Model.  The above 
five operators go under the name of the ``SUSY basis of 
operators'' in the literature \cite{gabbiani,damir}. 
 At intermediate stages of the matching calculation,
 we found it useful to introduce  matrix 
elements of two more operators,
\begin{eqnarray}
\label{ffop6}
Q6 &=& \left ( \psibar^i_b \gamma^\nu P_L \Psi^i_q \right ) 
 \left ( \psibar^j_b \gamma_\nu P_R \Psi^j_q \right ) \\
\label{ffop7}
Q7 &=& \left ( \psibar^i_b \gamma^\nu P_L \Psi^j_q \right ) 
 \left ( \psibar^j_b \gamma_\nu P_R \Psi^i_q \right ) .
\end{eqnarray}

Within perturbation theory matching can be carried out by considering 
scattering from an incoming state with  [heavy antiquark + light quark]
into an outgoing state with [heavy quark + light antiquark]. 
Symbolically, 
\be\label{inoutstates}
|in \rangle = |\overline{Q}^B ; q^C\rangle ,
\qquad 
\langle out | = \langle \overline{q}^A ; Q^D |,
\ee
where ``$A$'', ``$B$'', ``$C$'', and ``$D$'' are color indices. 
We also introduce external Dirac spinors $u_q$ and $v_q$ for 
the incoming light quark and outgoing light antiquark respectively, 
and similarly $\ubar$ and $\vbar$ for the outgoing heavy quark and 
incoming heavy antiquark. 

At the tree-level, matrix elements of 
operators $Q1$, $Q2$, $Q4$, and $Q6$ become,
\begin{eqnarray}
\label{tree1}
& &\langle out | \left ( \psibar^i_b \Gamma_1 \Psi^i_q \right ) 
 \left ( \psibar^j_b \Gamma_2 \Psi^j_q \right ) | in \rangle_{tree} 
=  \\ \nonumber
&+& \delta_{AB} \delta_{CD} \left [\, (\ubar \Gamma_1 u_q)
(\vbar \Gamma_2 v_q) + (\ubar \Gamma_2 u_q) (\vbar \Gamma_1 v_q) \, \right ] 
 \\ \nonumber
&-&\delta_{AD} \delta_{CB} \left [ \, (\ubar \Gamma_1 v_q)
(\vbar \Gamma_2 u_q) + (\ubar \Gamma_2 v_q) (\vbar \Gamma_1 u_q) \, \right ] .
\end{eqnarray}
Diagrammatically, we will denote the two Dirac structures , 
$S1 = (\ubar \Gamma_{1,2} u_q)
(\vbar \Gamma_{2,1} v_q) $ and
$S2 = (\ubar \Gamma_{1,2} v_q)
(\vbar \Gamma_{2,1} u_q) $ by the first and second diagrams in 
Fig.1 respectively.

\begin{figure}
\includegraphics[width=7.0cm,height=5.0cm]{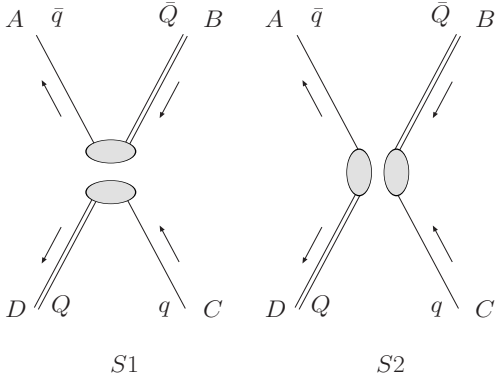}
\caption{ 
Diagrams depicting the two Dirac structures 
$S1 = (\ubar \Gamma_{1,2} u_q)
(\vbar \Gamma_{2,1} v_q) $ and 
$S2 = (\ubar \Gamma_{1,2} v_q)
(\vbar \Gamma_{2,1} u_q) $. The incoming and outgoing states are those of 
eq.(\ref{inoutstates}).  "A", "B", "C", and "D" are color indices.
 }
\end{figure}

\noindent
For operators $Q3$, $Q5$ and $Q7$ one has similarly,
\begin{eqnarray}
\label{tree2}
& &\langle out | \left ( \psibar^i_b \Gamma_1 \Psi^j_q \right ) 
 \left ( \psibar^j_b \Gamma_2 \Psi^i_q \right ) | in \rangle_{tree} 
=  \\ \nonumber
&-& \delta_{AB} \delta_{CD} \left [\, (\ubar \Gamma_1 v_q)
(\vbar \Gamma_2 u_q) + (\ubar \Gamma_2 v_q) (\vbar \Gamma_1 u_q) \, \right  ] 
 \\ \nonumber
&+&\delta_{AD} \delta_{CB} \left [ \, (\ubar \Gamma_1 u_q)
(\vbar \Gamma_2 v_q) + (\ubar \Gamma_2 u_q) (\vbar \Gamma_1 v_q) \, \right ] .
\end{eqnarray}
For the cases where $\Gamma_1 = \Gamma_2$ the two terms in the square 
brackets in eqns.(\ref{tree1}) and (\ref{tree2}) are identical and 
can be combined with a factor of 2.  When $\Gamma_1 \neq \Gamma_2$
 these terms should be kept 
separate during the matching procedure. The RHS of these equations 
can often be simplified using Fierz relations. In particular, using formulas 
of Appendix A one finds,
\be
\label{q67fierz}
\langle Q6 \rangle_{tree} = - 2 \, \langle Q5 \rangle_{tree}, 
\;\; \;\langle Q7 \rangle_{tree} = - 2 \, \langle Q4 \rangle_{tree}\, ,
\ee
and hence $Q6$ and $Q7$ are not independent operators.  Nevertheless, 
we have found the bookkeeping to be simplified if one first projects
onto $\langle Q6 \rangle_{tree}$ and $\langle Q7 \rangle_{tree}$ 
and then uses eq.(\ref{q67fierz}) to relate back to $Q5$ and $Q4$.

\begin{figure}
\includegraphics[width=8.5cm,height=7.0cm]{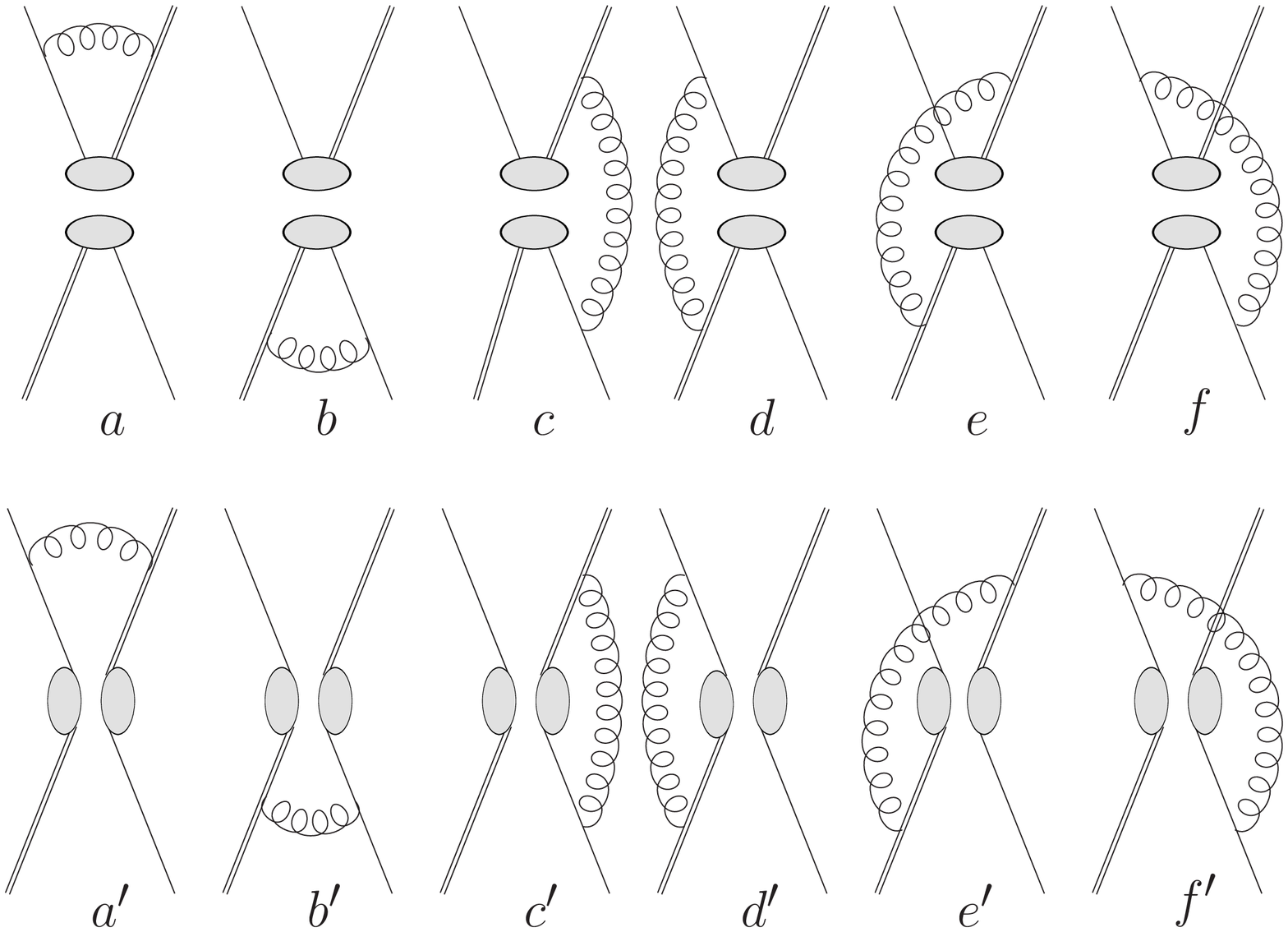}
\caption{ 
One-loop corrections to four fermion operators $Qk$.  The four external lines 
are the same as in Fig.1.
 }
\end{figure}

Continuum one-loop corrections to tree-level matrix elements are obtained by 
evaluating the diagrams in Fig.2a - Fig.2f$^\prime$.
 We carry out these calculations 
in the $\overline{MS}$-NDR scheme with definitions of general dimension 
four-fermion operators (i.e. of ``evanescent'' operators) as given in 
Appendix B \cite{bw,hn}. 
  We retain only terms of ${\cal O}(\alpha_s)$ and 
discard ${\cal O}(\alpha_s \Lambda_{QCD} / M)$ contributions, where 
M is the mass of the $b$ quark. We thus   
expand the full QCD calculations to the same order to which our effective 
theory (lattice NRQCD) perturbative calculations of the next section 
 are carried out. In this limit the heavy 
quark spinors obey,
\be
\label{hspinor}
\ubar \gamma_0 = \ubar, \qquad \vbar \gamma_0 = - \vbar.
\ee
As is well known, there is 
mixing among the four-fermion operators at one-loop.  For instance,
\begin{eqnarray}
\label{q1msbar}
& &\langle Q1 \rangle^{\overline{MS}} =   
\langle Q1 \rangle_{tree} \\ \nonumber
&&  + \alpha_s \, \left[\, c_{11}\,
\langle Q1 \rangle^{(0)}_{tree}  +   c_{12} \,
\langle Q2 \rangle^{(0)}_{tree}\, \right].
\end{eqnarray}
All matrix elements are taken  between $\langle out |$ and 
$| in \rangle$ 
 and the superscript $(0)$ means that we are working with spinors obeying 
(\ref{hspinor}).
Similarly, one finds that 
$\langle Q2 \rangle^{\overline{MS}}$ has contributions from $Q2$ and $Q1$, 
$\langle Q3 \rangle^{\overline{MS}}$  from $Q3$ and $Q1$, 
$\langle Q4 \rangle^{\overline{MS}}$  from $Q4$ and $Q6$, and
$\langle Q5 \rangle^{\overline{MS}}$  from $Q5$ and $Q7$.
The one-loop coefficients $c_{xy}$ depend on 
the $b$ quark mass $M$, the gluon mass $\lambda$ that acts as 
an IR regulator, and the $\overline{MS}$-NDR renormalization 
scale $\mu$. Their values for the operators in the basis (\ref{ffop1}) 
are given by,
\begin{eqnarray}
c_{11} &=& \frac{1}{4 \pi} \left \{ - \frac{35}{3} 
 -2 \,{\rm ln} \frac{\mu^2}{M^2} - 4 
\,{\rm ln} \frac{\lambda^2}{M^2}  \right \} \\
c_{12} &=& \frac{-8}{4 \pi}
\end{eqnarray}

\begin{eqnarray}
\label{c22}
c_{22} &=& \frac{1}{4 \pi} \left \{ 10  + 
\frac{16}{3} \,{\rm ln} \frac{\mu^2}{M^2} 
 - \frac{4}{3} \,{\rm ln} \frac{\lambda^2}{M^2}  \right \} \\
\label{c21}
c_{21} &=& \frac{1}{4 \pi} \left \{ \frac{3}{2}  + 
\frac{1}{3} \,{\rm ln} \frac{\mu^2}{M^2} 
 + \frac{2}{3} \,{\rm ln} \frac{\lambda^2}{M^2}  \right \} 
\end{eqnarray}

\begin{eqnarray}
\label{c33}
c_{33} &=& \frac{1}{4 \pi} \left \{ -2  - 
\frac{8}{3} \,{\rm ln} \frac{\mu^2}{M^2} 
 - \frac{4}{3} \,{\rm ln} \frac{\lambda^2}{M^2}  \right \} \\
\label{c31}
c_{31} &=& \frac{1}{4 \pi} \left \{ 3  + 
\frac{4}{3} \,{\rm ln} \frac{\mu^2}{M^2} 
 + \frac{2}{3} \,{\rm ln} \frac{\lambda^2}{M^2}  \right \} 
\end{eqnarray}

\begin{eqnarray}
c_{44} &=& \frac{1}{4 \pi} \left \{ \frac{143}{12}  + 
8 \,{\rm ln} \frac{\mu^2}{M^2} 
 - \frac{7}{2} \,{\rm ln} \frac{\lambda^2}{M^2}  \right \} \\
c_{46} &=& \frac{1}{4 \pi} \left \{ \frac{23}{8}  + 
  \frac{3}{4} \,{\rm ln} \frac{\lambda^2}{M^2}  \right \} 
\end{eqnarray}

\begin{eqnarray}
c_{55} &=& \frac{1}{4 \pi} \left \{ - \frac{85}{12}  -
 \,{\rm ln} \frac{\mu^2}{M^2} 
 - \frac{7}{2} \,{\rm ln} \frac{\lambda^2}{M^2}  \right \} \\
c_{57} &=& \frac{1}{4 \pi} \left \{- \frac{13}{8}  -
\frac{3}{2} \,{\rm ln} \frac{\mu^2}{M^2} 
+  \frac{3}{4} \,{\rm ln} \frac{\lambda^2}{M^2}  \right \} .
\end{eqnarray}
The one-loop coefficients $c_{11}$, $c_{12}$, $c_{22}$ 
and $c_{21}$ agree with the values already published 
in reference \cite{hashimoto}. Using eq.(\ref{q67fierz}) 
$c_{46}$ and $c_{57}$ can be replaced by,
\begin{eqnarray}
c_{45} &=& \frac{1}{4 \pi} \left \{- \frac{23}{4}  - 
  \frac{3}{2} \,{\rm ln} \frac{\lambda^2}{M^2}  \right \}  \\
c_{54} &=& \frac{1}{4 \pi} \left \{ \frac{13}{4}  +
3 \,{\rm ln} \frac{\mu^2}{M^2} 
-  \frac{3}{2} \,{\rm ln} \frac{\lambda^2}{M^2}  \right \} .
\end{eqnarray}


\section{ Matrix elements in the Effective Theory}
 In effective theories such as HQET or NRQCD one works 
separately with heavy quark fields that create heavy quarks ($\psibar_Q$) and
with those that 
annihilate heavy antiquarks ($\psibar_{\overline{Q}}$).  
At lowest  order  in $1/M$,  matrix elements
such as (\ref{tree1}) and (\ref{tree2}), are reproduced by working 
in the effective theory with 
\be
\label{effm0}
\hat{O} = \left ( \psibar_Q \Gamma_1 \Psi_q \right ) 
 \left ( \psibar_{\overline{Q}} \Gamma_2 \Psi_q \right ) 
+ \left ( \psibar_{\overline{Q}} \Gamma_1 \Psi_q \right ) 
 \left ( \psibar_Q \Gamma_2 \Psi_q \right ) .
\ee
If one introduces an effective theory field,
\be
\psibar^{eff}_b = \psibar_Q + \psibar_{\overline{Q}}
\ee
then $\psibar^{eff}_b$ and the QCD field $\psibar_b$ are related 
by a Foldy-Wouthuysen-Tani transformation.  In particular,
\be 
\label{fwt}
\psibar_b = \psibar^{eff}_b \, \left [ I + \frac{1}{2M} \vec{\gamma} \cdot 
\vec{\nabla}  \; + \; {\cal O}(1/M^2) \right ]\, ,
\ee
where the $\vec{\nabla}$ acts to the left.
By inserting (\ref{fwt}) into the expressions for the four-fermion 
operators $Q1$ - $Q7$, one sees that ${\cal O}(\Lambda_{QCD}/M)$ 
corrections to (\ref{effm0}) can be obtained within the 
effective theory by adding the following $1/M$ operator corrections
\begin{eqnarray}
\label{effj1}
\hat{O}j1 &=& \frac{1}{2M} \left [ \left ( 
 \vec{\nabla} \psibar_Q  \, \cdot \, \vec{\gamma} \, \Gamma_1 \Psi_q \right ) 
 \left ( \psibar_{\overline{Q}} \Gamma_2 \Psi_q \right ) \right . \\ \nonumber 
 &+& \left . \left ( 
 \psibar_Q  \Gamma_1 \Psi_q \right ) 
 \left ( \vec{\nabla}\psibar_{\overline{Q}}
\, \cdot \, \vec{\gamma} \, \Gamma_2 \Psi_q \right ) \right ] \; + \; 
(\Gamma_1 \rightleftharpoons \Gamma_2) .
\end{eqnarray}

\vspace{.1in}
\noindent
 As mentioned already, our effective theory consists of 
lattice NRQCD for heavy quarks, the improved staggered AsqTad for light quark 
action and Symanzik improved glue action. Details about these actions  
and their Feynman rules can be found for instance in reference \cite{pert1}. 
In this effective theory, we have evaluated 
one-loop corrections to matrix elements of $\hat{O}$ and $\hat{O}j1$. 
The one-loop corrections to $\hat{O}$ involve the same diagrams 
Fig.2a - 2f$^\prime$ as in continuum QCD. One obtains
\be
\label{cxyl}
\label{oeff}
\langle Q1 \rangle^{eff} = [ 1 + \alpha_s \, c^L_{11}]
\langle Q1 \rangle^{(0)}_{tree} \; + \; \alpha_s \, c^L_{12} 
\langle Q2 \rangle^{(0)}_{tree} \, ,
\ee
and similarly for all the other $Qk$'s. 
Several of the lattice one-loop integrals are IR divergent and we use 
a gluon mass $\lambda$ to extract the IR finite contributions 
as explained in reference \cite{pert1}. 
The $c^L_{xy}$'s have the same IR divergent ${\rm ln}(\lambda^2)$ terms 
as the corresponding $c_{xy}$ in continuum QCD, and they also 
depend on the bare heavy quark mass. The divergent terms 
will cancel when we do the matching (see below) and things are 
reduced to finite differences such as $[c_{xy} - c^L_{xy}]$. 
We carry out 
the lattice perturbative calculations in both Feynman and Landau gauges 
and use gauge invariance as a check on our results.

\vspace{.1in}
\noindent
In order to calculate the one-loop renormalization coefficients for 
the matrix elements of the operators $\hat{O}j1$ the diagrams 
of Figs.3 \& 4 need to be evaluated.  One finds, 
\be
\label{zetaxy}
\label{oj1eff}
\langle Q1j1 \rangle^{eff} = \langle Q1j1 \rangle^{(0)}_{tree} + 
\alpha_s \, \left[\,\zeta^{11} \langle Q1 \rangle^{(0)}_{tree} +
  \zeta^{12} \langle Q2 \rangle^{(0)}_{tree} \, \right ]\, ,
\ee
where we ignore ${\cal O}(\alpha_s)$ corrections to $\langle Q1j1 \rangle 
^{(0)}_{tree}$.  Similar expressions are obtained for the other 
$\hat{O}j1$.  

The coefficients $\zeta^{xy}$ tell us about the 
``mixing down'' of a dimension seven operator $\hat{O}j1$ onto 
dimension six operators $\hat{O}$.  On dimensional grounds these coefficients 
go as $1/(aM)$, ``$a$'' being the lattice spacing.  They represent 
``power law'' contributions from matrix elements in the effective theory. 
Power law terms are unavoidable when working with effective theories and 
they need to be subtracted in order that the 
effective theory produce the same physics as full QCD (which does not 
suffer from power law contributions).  Our matching procedure, described 
in the next section, will be such that power law contributions are 
removed from matrix elements of $\hat{O}j1$ through ${\cal O}(\alpha_s/
(aM))$. Errors from the mismatch between QCD and effective theory due to 
power law contributions will come in at ${\cal O}(\alpha_s^2/(aM))$.

\begin{figure}
\includegraphics[width=8.0cm,height=5.0cm]{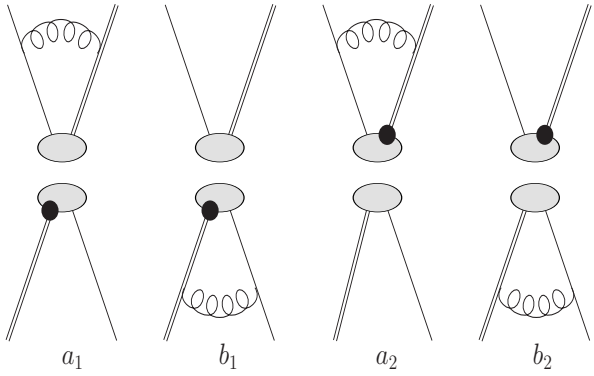}
\caption{ 
One loop corrections to the $1/M$ corrections $\hat{O}j1$.  The dark dot 
denotes a derivative acting on either the heavy quark or the 
heavy anti-quark propagator.  We only show corrections 
associated with Fig.2a and 2b. Similar diagrams exist 
for corrections to 2c -- 2f$^\prime$.
 }
\end{figure}

\begin{figure}
\includegraphics[width=7.0cm,height=7.0cm]{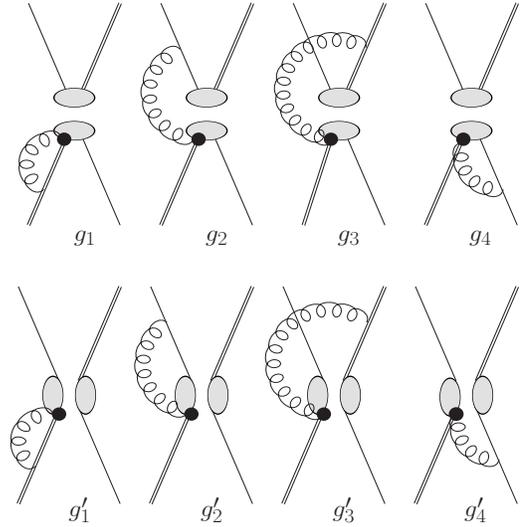}
\caption{ 
Further possible diagrams associated with $1/M$ corrections $\hat{O}j1$.  
In practice, however, these do not contribute to the $\zeta^{xy}$. 
Diagrams where the derivative acts on the heavy anti-quark propagator 
(top right line) similarly do not contribute. 
 }
\end{figure}


\section{ Matching }
We wish to relate the continuum QCD matrix elements
 $\langle Qk \rangle^{\overline{MS}}$ to the matrix elements 
$\langle Qk \rangle^{eff}$ in the effective theory.  The latter will 
ultimately be replaced by output from nonperturbative simulations. We will
focus on matching of $Q1$.  The other $Qk$'s are handled 
identically. 

If one expands the first term on the RHS of (\ref{q1msbar}) in 
powers of $1/M$, i.e. $\langle Q1 \rangle_{tree} \rightarrow 
\langle Q1 \rangle^{(0)}_{tree} + \langle Q1j1 \rangle^{(0)}_{tree}$, 
then this equation becomes
\begin{eqnarray}
\label{q1msbar2}
\langle Q1 \rangle^{\overline{MS}} &=& [ 1 + \alpha_s \, c_{11}]
\langle Q1 \rangle^{(0)}_{tree} \\ \nonumber
& + & \alpha_s \, c_{12} 
\langle Q2 \rangle^{(0)}_{tree} 
\; + \; \langle Q1j1\rangle^{(0)}_{tree}.
\end{eqnarray}
The next step is to rewrite the matrix elements $\langle ... \rangle
^{(0)}_{tree}$ appearing on the RHS in terms of matrix elements in the 
effective theory $\langle ... \rangle^{eff}$.  This can be accomplished 
by inverting (\ref{oeff}) and (\ref{oj1eff}). To the order that we are working
one has,
\be
\label{oeff2}
\langle Q1 \rangle^{(0)}_{tree} = \langle Q1 \rangle^{eff}
- \alpha_s \, \left [\, c^L_{11}] \,
\langle Q1 \rangle^{eff} \, + \,  c^L_{12} \,
\langle Q2 \rangle^{eff} \, \right ] ,
\ee
\be 
\alpha_s \, \langle Q2 \rangle^{(0)}_{tree} \Rightarrow \alpha_s 
\langle Q2 \rangle^{eff} ,
\ee
and
\be
\label{oj1eff2}
\langle Q1j1 \rangle^{(0)}_{tree} = 
\langle Q1j1 \rangle^{eff}
- \alpha_s \, \left [\, \zeta^{11} \,
\langle Q1 \rangle^{eff} \, + \,  \zeta^{12} \,
\langle Q2 \rangle^{eff} \, \right ] .
\ee
Upon inserting the last three equations into (\ref{q1msbar2}) one 
ends up with,
\begin{eqnarray}
\label{q1msbar3}
&&\langle Q1 \rangle^{\overline{MS}} = \\ \nonumber
&& [ \, 1 + \alpha_s \, \rho_{11} \,]
 \, \langle Q1 \rangle^{eff} 
  \, + \, \alpha_s \, \rho_{12} \, \langle Q2 \rangle^{eff} +  \\ \nonumber
&  & \langle Q1j1 \rangle^{eff}
- \alpha_s \, \left [\, \zeta^{11} \,
\langle Q1 \rangle^{eff} \, + \,  \zeta^{12} \,
\langle Q2 \rangle^{eff} \, \right ] \\ \nonumber
&& \; \; + \;{\cal O}(\alpha_s^2, \alpha_s \Lambda_{QCD} / M),
\end{eqnarray}
where,
\be \label{rhodef}
 \rho_{xy} = c_{xy} - c^L_{xy} .
\ee
With (\ref{q1msbar3}) we have achieved the goal of relating the continuum 
full QCD matrix element $\langle Q1 \rangle ^{\overline{MS}}$ to
 matrix elements in the effective theory.  Although this perturbative 
matching calculation was carried out with external scattering states, one 
carries over matchings such as (\ref{q1msbar3}) to the case of hadronic matrix 
elements between  $B_q^0$ and $\overline{B^0}_q$ states where then 
$\langle Q1 \rangle^{eff}$ or $\langle Q1j1 \rangle^{eff}$ must 
be evaluated nonperturbatively. With such nonperturbative 
matrix elements in mind we define  
\begin{eqnarray}
 && \langle Q1j1 \rangle^{eff} 
- \alpha_s \, \left [\, \zeta^{11} \,
\langle Q1 \rangle^{eff} \, + \,  \zeta^{12} \,
\langle Q2 \rangle^{eff} \, \right ] \\ \nonumber
&&  \equiv \langle Q1j1 
\rangle ^{eff}_{sub}.
\end{eqnarray}
The combination $\langle Q1j1 \rangle^{eff}_{sub}$ represents thus 
the matrix element of the 
dimension seven $1/M $ correction in the effective theory with power law 
contributions subtracted out through ${\cal O}(\alpha_s/(aM))$. 
This matrix element gives us the physical 
$\Lambda_{QCD}/M$ contributions to $\langle Q1 \rangle^{\overline{MS}}$ up 
to corrections of ${\cal O}(\frac{\alpha_s^2}{(aM)})$. 
Further discussion of power law subtractions for the case of 
heavy-light currents are given in reference \cite{collins}. 
  A more complete derivation of one-loop 
matching formulas including all contributions at ${\cal O}(\alpha_s 
\Lambda_{QCD}/M)$ 
is provided (again for heavy-light currents) in 
reference \cite{pert2}.

A final technical detail is that eq.(\ref{q1msbar3}) differs from 
eq.(10) of \cite{ourbsmix} in that here we assume the same 
normalization of states in the effective theory as in continuum full 
QCD.  For the purposes of evaluating matching coefficients it 
is convenient to do so. Any differences in normalization of states are 
taken care of at the stage of doing the
non-perturbative calculations 
 and of extracting  decay constants and bag parameters.


\section{Results}
In this section we summarize results for the effective theory coefficients 
$c^L_{xy}$ and $\zeta^{xy}$  of eqns.(\ref{cxyl}) and (\ref{zetaxy}), 
and for the matching coefficients $\rho_{xy}$ of eq.(\ref{q1msbar3}). 
We present numbers for three values of the bare heavy quark mass
in lattice units, $aM_0$,
corresponding to the $b$ quark mass on lattices with spacings 
$0.09fm$, $0.12fm$ and $0.17fm$, as fixed in previous studies of the 
$\Upsilon$ system \cite{agray}. 
These values of $a$ correspond to the so-called MILC fine, 
coarse and super-coarse lattices,  
which have been used extensively in recent studies of heavy-heavy 
\cite{agray} and heavy-light \cite{fB,formfactor} 
quantities with the same choice of lattice actions as 
the one in the present article. 

In Table I we list values for the one-loop renormalization coefficients 
$c^L_{xy}$ after subtracting the IR divergent  
${\rm ln}(a \lambda)^2$ pieces, together with the one-loop $\zeta^{xy}$, 
which are IR finite. Table II shows values for $\rho_{xy}$ at scale 
$\mu$ equal to the heavy quark mass $M$. The parameter n in these two 
tables is the stability parameter of the NRQCD action.
\begin{table}
\begin{center}
\begin{tabular}{|c|c|c|c|}
\hline
$aM_0$ (n) & $1.95$ $(n=2)$& $2.8$ $(n=2)$ & $ 4.0$ $(n=2)$ \\
\hline
$c^L_{11}$  & $-1.196$ & $-0.735$ & $-0.403$  \\
$c^L_{12}$  & $-1.802$ & $-1.315$ & $-0.960$ \\
$c^L_{22}$  & $\;\;0.010$ & $\;\; 0.014$ & $-0.004$ \\
$c^L_{21}$  & $\;\;0.020$ & $-0.018$ & $-0.050$ \\
$c^L_{33}$  & $-0.890$& $-0.644$ & $-0.483$ \\
$c^L_{31}$  & $\;\;0.133$ & $\;\; 0.064$ & $\;\;0.010$  \\
$c^L_{44}$  & $\;\;0.692$ & $\;\; 0.599$ & $\;\;0.529$ \\
$c^L_{46}$  & $\;\;0.097$ & $\;\; 0.029$ & $-0.022$ \\
$c^L_{55}$  & $-0.886$ & $-0.553$ & $-0.311$ \\
$c^L_{57}$  & $-0.467$ & $-0.383$ & $-0.322$ \\
$c^L_{45}$  & $-0.194$ & $-0.058$ & $\;\;0.044$ \\
$c^L_{54}$  & $\;\;0.934$ & $\;\;0.766$ & $\;\;0.644$\\
\hline
$\zeta^{11}$  & $\;\;0.218$ & $\;\; 0.166$ & $\;\;0.123$  \\
$\zeta^{12}$  & $\;\;0.874$ & $\;\; 0.662$ & $\;\;0.492$ \\
$\zeta^{22}$  & $\;\;0.364$ & $\;\; 0.276$ & $ \;\;0.206$ \\
$\zeta^{21}$  & $\;\;0.009$ & $\;\; 0.007$ & $\;\; 0.005$ \\
$\zeta^{33}$  & $-0.073$ & $-0.055$ &  $-0.042$\\
$\zeta^{31}$  & $ \;\;0.064$ & $\;\; 0.048$ & $\;\;0.035$  \\
$\zeta^{44}$  & $\;\;0.309$ & $\;\;0.235$ & $\;\;0.175$ \\
$\zeta^{46}$  & $-0.046$ & $-0.034$ & $-0.026$ \\
$\zeta^{55}$  & $ \;\;0.091$ & $\;\;0.069$ & $\;\;0.051$ \\
$\zeta^{57}$  & $ \;\;0.064$ & $ \;\;0.048$ & $\;\;0.035$ \\
$\zeta^{45}$  & $ \;\;0.092$ & $ \;\;0.068$ & $\;\;0.052$ \\
$\zeta^{54}$  & $-0.128$ & $-0.096$ & $-0.070$ \\
\hline
\end{tabular}
\caption{ 
The one-loop coefficients $c^L_{xy}$ and $\zeta^{xy}$ for 
three values of $aM_0$. $n$ is the stability parameter 
in the NRQCD action. 
 IR divergent ${\rm ln}(a\lambda)^2$ terms 
are omitted.  Numerical integration errors are of order one or less 
in the last digit.
}
\end{center}
\end{table}

\begin{table}
\begin{center}
\begin{tabular}{|c|c|c|c|}
\hline
$aM_0$ (n) & $1.95$ $(n=2)$& $2.8$ $(n=2)$ & $ 4.0$ $(n=2)$ \\
\hline
$\rho_{11}$  & $\;\;0.693$ & $\;\;0.462$ &  $\;\;0.357$\\
$\rho_{12}$  & $\;\;1.165$ & $\;\;0.678$ & $\;\;0.323$ \\
$\rho_{22}$  & $\;\;0.927$& $\;\;1.000$ & $\;\;1.094$ \\
$\rho_{21}$  & $\;\;0.029$ & $\;\;0.028$ & $\;\;0.022$ \\
$\rho_{33}$  & $\;\;0.873$ & $\;\;0.703$ & $\;\;0.618$  \\
$\rho_{31}$  & $\;\;0.035$ & $\;\;0.065$ & $\;\;0.082$  \\
$\rho_{44}$  & $\;\;0.628$ & $\;\;0.923$ & $\;\;1.192$ \\
$\rho_{46}$  & $\;\;0.052$ & $\;\;0.077$ & $\;\;0.085$  \\
$\rho_{55}$  & $\;\;0.694$ & $\;\;0.563$ & $\;\;0.520$  \\
$\rho_{57}$  & $\;\;0.258$ & $\;\;0.131$ & $\;\;0.027$  \\
$\rho_{45}$  & $-0.104$ & $-0.154$ & $-0.170$  \\
$\rho_{54}$  & $-0.516$ & $-0.262$ & $-0.054$  \\
\hline
\end{tabular}
\caption{ 
The one-loop matching coefficients $\rho_{xy}$ in (\ref{rhodef})
with $\mu \equiv M$ for three values of $aM_0$.  
}
\end{center}
\end{table}

\begin{table}[b]
\begin{center}
\begin{tabular}{|c|c|c|c|c|}
\hline
   & \multicolumn{4}{c|} {$\xi = 1 \qquad \qquad \qquad \qquad \xi = 0$} \\
\hline
diagram & &  $\frac{1}{4 \pi} {\rm ln}(a \lambda)^2 \times$ &
 &  $\frac{1}{4 \pi} {\rm ln}(a \lambda)^2 \times$ \\
\hline
\hline
$a + b$ & $0.836 \times 2$  & $-\frac{4}{3} \times 2$ & $0.3274 \times 2$ & 0 \\
$a^\prime + b^\prime$ & $0.0823 \times 2$ & 0 & $0.0823 \times 2$ & 0 \\
$c + d$  & $ -0.0495 \times 2$ & $\frac{1}{6} \times 2$ & $ 0.0139 \times 2$ & 0 \\
$e$ & $0.0331 $ & $\frac{1}{3} $ & $ -0.0304 $ & $ \frac{1}{2}$ \\
$ e^\prime$ & $ 0.0269$   & 0 & $ 0.0269$  & 0 \\
$f$ & $0.0640 $ & $- \frac{1}{6} $ & $0.0006 $  & 0 \\
\hline
$Z_q$ & $ -0.924$  & $\frac{4}{3}$  & $ -0.416 $ & 0 \\
$Z_Q$ & $-0.338 $ & $ - \frac{8}{3}$ & $ 0.171$ & $-4$ \\
\hline
\hline
{\bf Total }
   & $0.5996$& $ -\frac{7}{2}$& $0.5993$&  $ - \frac{7}{2}$\\
\hline
\end{tabular}
\caption{ 
Contributions to the coefficient $c^L_{44}$ from the diagrams of Fig.2
for $aM_0= 2.8$. $\xi = 1$ and $\xi = 0$ refer to Feynman or Landau 
gauge respectively.  
 The second and fourth 
columns give the IR finite contributions.  Columns 3 and 5 list IR divergent 
terms in units of $\frac{1}{4 \pi} {\rm ln}(a \lambda)^2$. $Z_q$ and $Z_Q$ 
are the light and heavy quark wave function renormalizations respectively 
and are taken from reference \cite{pert1}. 
The last row gives the full $c^L_{44}$.
}
\end{center}
\end{table}

\begin{table}
\begin{center}
\begin{tabular}{|c|c|c|c|c|}
\hline
   & \multicolumn{4}{c|} {$\xi = 1 \qquad \qquad \qquad \qquad \xi = 0$} \\
\hline
diagram & &  $\frac{1}{4 \pi} {\rm ln}(a \lambda)^2 \times$ &
 &  $\frac{1}{4 \pi} {\rm ln}(a \lambda)^2 \times$ \\
\hline
\hline
$a^\prime + b^\prime$ & $-0.0743 \times 2$  & $\frac{1}{4} \times 2$ &
 $0.0209 \times 2$ & 0 \\
$c + d$  & $ 0.0137 \times 2$ & 0  & $0.0137 \times 2$ & 0 \\
$e$ & $0.0045 $ & 0  & $ 0.0045 $ & 0 \\
$ e^\prime$ & $ 0.0497$   & $ \frac{1}{2}$ & $ -0.0457$  & $ \frac{3}{4}$ \\
$f^\prime$ & $0.0960 $ & $- \frac{1}{4} $ & $0.0009 $  & 0 \\
\hline
\hline
{\bf Total }&$ 0.029$& $ \frac{3}{4}$& $0.029$&  $  \frac{3}{4}$\\
\hline
\end{tabular}
\caption{ 
Same as Table III for the coefficient $c^L_{46}$.
}
\end{center}
\end{table}

In Table III we illustrate how the different diagrams contribute 
to $c^L_{44}$ for $aM_0 = 2.8$.
  Calculations were done in both Feynman ($\xi = 1$) and Landau 
($\xi = 0$) gauges to check for gauge invariance. We have evaluated 
$c^L_{44}$ by collecting all contributions that are 
proportional to $\delta_{AB} \delta_{CD}$ and that can be written 
 (using Fierz 
relations to convert where necessary) in terms of Dirac structures 
$  (\ubar P_L u_q)
(\vbar P_R v_q)$ or $ (\ubar P_R u_q) (\vbar P_L v_q) $. 
 We could just as well have collected terms 
proportional to $- \delta_{AD} \delta_{CB}$ and of Dirac structure 
$  (\ubar P_L v_q)
(\vbar P_R u_q)$ or $ (\ubar P_R v_q) (\vbar P_L u_q) $ 
to obtain the same final result.
Table IV illustrates  different contributions 
to $c^L_{46}$ again for $aM_0 = 2.8$.  Here we collect terms 
proportional to $\delta_{AB} \delta_{CD}$ and of Dirac structure 
$  (\ubar \gamma^\nu P_L u_q)
(\vbar \gamma_\nu P_R v_q)$ or $ (\ubar \gamma^\nu P_R u_q)
 (\vbar \gamma_\nu P_L v_q) $. From $c^L_{46}$ one easily 
obtains $c^L_{45} = -2\,c^L_{46}$.


\section{Summary}

We have completed the one-loop matching of a complete set of $\Delta B = 2$ 
heavy-light four fermion operators through ${\cal O}(\alpha_s, \Lambda_{QCD}/
M_b, \alpha_s/(aM_b))$.  The main results are the coefficients $\rho_{xy}$ of 
Table II and the $\zeta^{xy}$ of Table I.  

We find that with the lattice actions employed in this article 
(and in our simulations) matching coefficients are all well behaved. 
None of them are particularly large, and in fact many are considerably 
smaller than one. 
An interesting feature that can be extracted from the results 
in Table III and IV, and which holds also for the matrix elements of  
the other operators, is that the one-loop matching coefficients are 
dominated by the current-like diagrams, $a$ and $b$, and the wave function 
renormalizations. The contributions from the pure four-fermion 
diagrams are at least an order of magnitude smaller. We are currently 
investigating non-perturbative matching methods for 
heavy(NRQCD)-light(staggered) currents. The same methodology could 
be applied here for four-fermion operators to calculate 
non-perturbatively the main contribution to the renormalization 
coefficients. In this way we could considerably reduce the uncertainty 
associated with the matching process, which is one of the main 
sources of error at present in our calculation 
of $f_{B_s}\sqrt{B_{B_s}}$ and $f_{B_d}\sqrt{B_{B_d}}$.

The matching calculation in this article is an important part of 
the HPQCD collaboration's studies of  $B_s$ and $B_d$ meson mixing 
via lattice QCD methods both in the Standard Model 
\cite{ourbsmix,elvira,ourbdmix} and beyond. Values 
for the mass and decay width differences, $\Delta M_q$ and 
$\Delta \Gamma_q$ with $q=s,d$, as well as for the ratio 
$\xi=\frac{f_{B_s}\sqrt{B_{B_s}}}{f_{B_d}\sqrt{B_{B_d}}}$, which 
use the results presented here will be available soon \cite{ourbdmix}. 
Extensions of this work to matching with 
other lattice actions such as the HISQ light quark action are
straightforward and are planned for the future.


\vspace{.1in}
{\bf Acknowledgements}: \\
This work was supported by the DOE (USA), by NSERC (Canada) 
and by the Junta de Andaluc\'{\i}a [P05-FQM-437 and P06-TIC-02302] (Spain). 
The numerical integrals were carried out in part on facilities 
of the USQCD Collaboration, which are funded by the Office of 
Science of the U.S. Department of Energy. 
JS thanks the theory groups at TRIUMF and Simon Fraser University 
for support and hospitality while part of this project was carried out.  
She is also grateful to Glasgow University and SUPA for hospitality 
during later stages of this work. The authors thank A.Buras and U.Nierste 
for useful correspondence on evanescent operators.

\appendix

\section{Fierz Relations in 4 D}
In this appendix we collect some Fierz relations that were found to be 
useful in our calculations. They are written in terms of
fixed external  spinors rather 
than fermionic fields, which means that interchanging two spinors does not bring 
in a minus sign.  In our calculations, fermionic signs come in at the 
stage of doing the Wick contractions, for instance to get the RHS's of 
eq.(\ref{tree1}) and (\ref{tree2}).
\be
[\ubar \gamma^\nu P_L v_q ]\, [\vbar \gamma_\nu P_L u_q] = - [\ubar \gamma^\nu P_L 
u_q] \, [\vbar \gamma_\nu P_L v_q]
\ee
\begin{eqnarray}
\label{fz2}
&& [\ubar P_L v_q ]\, [\vbar P_L u_q] = \\ \nonumber
&& \frac{1}{2} \, [\ubar P_L u_q] \, [\vbar  P_L v_q] 
+ \frac{1}{8} \,
 [\ubar \sigma^{\mu \nu} P_L u_q] \, [\vbar \sigma_{\mu \nu}  P_L v_q] 
\end{eqnarray}
where $\sigma^{\mu \nu} = \frac{i}{2} [\gamma^\mu, \gamma^\nu]$.
\be
[\ubar P_L v_q ]\, [\vbar  P_R u_q] = \frac{1}{2} \, [\ubar \gamma^\nu P_R 
u_q] \, [\vbar \gamma_\nu P_L v_q]
\ee
\be
[\ubar \gamma^\nu P_L v_q ]\, [\vbar \gamma_\nu P_R u_q] = 2 \, [\ubar P_R 
u_q] \, [\vbar  P_L v_q]
\ee
In the large $M$ limit, i.e. with spinors obeying (\ref{hspinor}), 
the relation (\ref{fz2}) simplifies to \cite{flynn},
\begin{eqnarray}
&& [\ubar P_L v_q ]\, [\vbar P_L u_q] = \\ \nonumber
&& [\ubar P_L u_q] \, [\vbar  P_L v_q] 
+ \frac{1}{2} \,
 [\ubar \gamma^\nu P_L u_q] \, [\vbar \gamma_\nu  P_L v_q] .
\end{eqnarray}


\section{Some Formulas in General Dimensions}
We carry out the continuum one-loop calculations using dimensional 
regularization in the $\overline{MS}$ and NDR scheme.  As is by now well 
known, this information is insufficient to fix one's renormalization 
scheme unambiguously.  One needs to specify in addition how one handles 
d-dimensional Dirac structures 
appearing at intermediate stages of the calculations and how
they  are projected onto some 4-dimensional basis.  This is because 
the Dirac algebra is infinite dimensional for non-integer $d$, and 
requires in addition to the 4-d basis set an infinite set of 
``evanescent operators''.  The ${\cal O}(\epsilon)$ 
(we use $ d \equiv 4 - \epsilon$) terms in the projections 
onto the 4-d basis defines one's choice of evanescent operators, and 
is convention dependent \cite{bw,hn,bb,bm}.
In our calculations we have adopted the following 
conventions taken from the literature \cite{bb,bm}
\begin{eqnarray}
\label{b1}
 &&[\ubar \gamma^\alpha \gamma^\beta \gamma^\nu P_L u_q] \,
[\vbar \gamma_\alpha \gamma_\beta \gamma_\nu P_L v_q] \\ \nonumber
&& \qquad \Rightarrow (16 - 2 \epsilon)
[\ubar \gamma^\nu P_L u_q] \, [\vbar \gamma_\nu P_L v_q ]
\end{eqnarray}
\begin{eqnarray}
\label{b2}
&& [\ubar \gamma^\alpha \gamma^\beta \gamma^\nu P_L u_q] \,
[\vbar \gamma_\nu \gamma_\beta \gamma_\alpha P_L v_q] \\ \nonumber
&& \qquad \Rightarrow (4 - 4 \epsilon)
[\ubar \gamma^\nu P_L u_q] \, [\vbar \gamma_\nu P_L v_q ]
\end{eqnarray}
\begin{eqnarray}
\label{b3}
&& [\ubar \gamma^\alpha \gamma^\beta P_L u_q] \,
[\vbar \gamma_\alpha \gamma_\beta P_L v_q] \\ \nonumber
&& \qquad \Rightarrow (8 - 2 \epsilon)
[\ubar P_L u_q] \, [\vbar  P_L v_q ] \\ \nonumber
&& \qquad  - (8 - 4 \epsilon)
[\ubar P_L v_q] \, [\vbar  P_L u_q ]
\end{eqnarray}
\begin{eqnarray}
\label{b4}
&& [\ubar \gamma^\alpha \gamma^\beta  P_L u_q] \,
[\vbar \gamma_\alpha \gamma_\beta  P_R v_q]  \\ \nonumber
&& \qquad \Rightarrow (4 + 2 \epsilon)
[ \ubar P_L u_q ] \, [\vbar  P_R v_q ]
\end{eqnarray}
\begin{eqnarray}
\label{b5}
&& [\ubar \gamma^\alpha \gamma^\beta  P_L u_q] \,
[\vbar \gamma_\beta \gamma_\alpha  P_R v_q]  \\ \nonumber
&& \qquad \Rightarrow (4 - 4 \epsilon)
[ \ubar P_L u_q ] \, [\vbar  P_R v_q ]
\end{eqnarray}
In adopting (\ref{b3}) we  are following the conventions of 
reference \cite{bb}.

We note that \cite{bm} has  different 
conventions from \cite{bb}. Instead of eq.(\ref{b3}) the 
\cite{bm} conventions translate into the relation
\begin{eqnarray} \label{bmconvention}
&& [\ubar \gamma^\alpha \gamma^\beta P_L u_q] \,
[\vbar \gamma_\alpha \gamma_\beta P_L v_q] \\ \nonumber
&& \qquad \Rightarrow (8 - \epsilon)
[\ubar P_L u_q] \, [\vbar  P_L v_q ] \\ \nonumber
&& \qquad  - 8 \,
[\ubar P_L v_q] \, [\vbar  P_L u_q ]
\end{eqnarray}
Using (\ref{bmconvention}) leads to different constant terms in results 
for the one-loop coefficients $c_{22}$, $c_{21}$, $c_{33}$ and 
$c_{31}$.  The ${\rm ln}\mu^2$ and ${\rm ln}\lambda^2$ terms 
remain unchanged, however. Denoting by $\tilde{c}_{ij}$ the coefficients  
obtained by using the \cite{bm} conventions, one finds that eqns.(\ref{c22}), 
(\ref{c21}), (\ref{c33}) and (\ref{c31}) are modified to,
\begin{eqnarray}
\tilde{c}_{22} &=& \frac{1}{4 \pi} \left \{ 6  + 
\frac{16}{3} \,{\rm ln} \frac{\mu^2}{M^2} 
 - \frac{4}{3} \,{\rm ln} \frac{\lambda^2}{M^2}  \right \} \\
\tilde{c}_{21} &=& \frac{1}{4 \pi} \left \{ \frac{4}{3}  + 
\frac{1}{3} \,{\rm ln} \frac{\mu^2}{M^2} 
 + \frac{2}{3} \,{\rm ln} \frac{\lambda^2}{M^2}  \right \} 
\end{eqnarray}

\begin{eqnarray}
\tilde{c}_{33} &=& \frac{1}{4 \pi} \left \{  - 
\frac{8}{3} \,{\rm ln} \frac{\mu^2}{M^2} 
 - \frac{4}{3} \,{\rm ln} \frac{\lambda^2}{M^2}  \right \} \\
\tilde{c}_{31} &=& \frac{1}{4 \pi} \left \{ \frac{67}{12}  + 
\frac{4}{3} \,{\rm ln} \frac{\mu^2}{M^2} 
 + \frac{2}{3} \,{\rm ln} \frac{\lambda^2}{M^2}  \right \} .
\end{eqnarray} 
Care is required to ensure that a consistent set of conventions 
is applied to different parts of a calculation 
contributing to a physical 
quantity such as $\Delta M_q$.


\end{document}